\documentclass[useAMS,usenatbib]{mn2e}

\usepackage{graphicx}
\usepackage{amssymb}
\usepackage{lscape}

\def \oii {[O\,{\sc ii}]~}
\def \oiii {[O\,{\sc iii}]~}
\def \halpha {H$\alpha$}
\def \hbeta {H$\beta$}
\def \hi {H\,{\sc{i}}}

\def \kms {km\,s$^{-1}$}
\def \vsys {$v_{sys}$}
\def \w50 {$W_{50}$}

\title[Comparison of \hi\ and optical redshifts of galaxies]{Comparison of \hi\ and optical
  redshifts of galaxies - The impact of redshift uncertainties on
  spectral line stacking}

\author[]
{Natasha Maddox\thanks{nmaddox@ast.uct.ac.za}$^{1}$, Kelley M. Hess$^{1}$,
S.-L. Blyth$^1$, M.~J. Jarvis$^{2,3}$
\vspace*{6pt}\\
$^1$Astrophysics, Cosmology and Gravity Centre (ACGC), Astronomy
Department, University of Cape Town, \\
Private Bag X3, 7701 Rondebosch, Republic of South Africa\\
$^2$Oxford Astrophysics, Denys Wilkinson Building,
University of Oxford, Keble Rd, Oxford, OX1 3RH, UK\\
$^3$Physics Department, University of the Western Cape,
Cape Town, 7535, Republic of South Africa}

\begin{document}


\maketitle

\begin{abstract}

Accurate optical redshifts will be critical for spectral co-adding
techniques used to extract detections from below the noise level in
ongoing and upcoming surveys for \hi, which will extend our current
understanding of gas reservoirs in galaxies to lower column densities
and higher redshifts. We have used existing, high quality optical and
radio data from the SDSS and ALFALFA surveys to investigate the
relationship between redshifts derived from optical spectroscopy and
neutral hydrogen (\hi) spectral line observations. We find that the
two redshift measurements agree well, with a negligible systematic
offset and a small distribution width. Employing simple simulations,
we determine how the width of an ideal stacked \hi\ profile depends on
these redshift offsets, as well as larger redshift errors more
appropriate for high redshift galaxy surveys. The width of the stacked
profile is dominated by the width distribution of the input individual
profiles when the redshift errors are less than the median width of
the input profiles, and only when the redshift errors become large,
$\sim$150\,\kms, do they significantly affect the width of the stacked
profile. This redshift accuracy can be achieved with moderate
resolution optical spectra. We provide guidelines for the number of
spectra required for stacking to reach a specified mass sensitivity,
given telescope and survey parameters, which will be useful for
planning optical spectroscopy observing campaigns to supplement the
radio data. 

\end{abstract}


\section{Introduction}\label{sec:Introduction}

Measurements of the neutral hydrogen (\hi) gas content of galaxies as
a function of redshift are critical to our understanding of galaxy
formation and evolution. In the local Universe, large area surveys
such as the \hi\ Parkes All-Sky Survey (HIPASS, \citealt{Barnes2001},
\citealt{Meyer2004}) and the Arecibo Legacy Fast ALFA survey (ALFALFA,
\citealt{Giovanelli2005a}) have evaluated metrics such as the \hi\
mass function and the local \hi\ matter density ($\Omega_{HI}$) with
increasing accuracy (see, for example \citealt{Zwaan2005},
\citealt{Martin2010}). Combined with complementary multi-wavelength
photometry and spectroscopy, these surveys have greatly expanded our
understanding of the gaseous and stellar components of local galaxies
as a function of environment, and provide a $z=0$ reference for
cosmological simulations that incorporate gas physics into galaxy formation.

A galaxy's \hi\ gas content serves as a reservoir of fuel for future
star formation.  Surveys of the stellar content of galaxies show that 
the star formation rate decreases by an order of magnitude from
$z\sim1$ to the present day (\citealt{Madau1998}, \citealt{Hopkins2006}, for
example), however, little is known of the \hi\ content of galaxies in
this epoch.  Simulations by \citet{Obreschkow2009} and
\citet{Lagos2011} predict the evolution in the size, velocity
profile, Tully-Fisher relation, and \hi\ mass function of \hi\ disks
as a function of redshift.  However, the bandwidth accessible to
existing radio telescopes, combined with the intrinsically weak signal from the 21 cm
line of neutral hydrogen, have made comprehensive surveys of \hi\
emission at redshifts beyond $z>0.1$ technically challenging, or
prohibitively expensive in terms of telescope time.

Observations seeking to detect \hi\ in emission at intermediate
redshift have employed a number of different strategies.
\citet{Catinella2008} used the Arecibo Observatory to detect 10 massive
field galaxies out to a redshift of $z=0.25$. While single dish
telescopes are very sensitive, they lack the spatial resolution to
resolve individual galaxies at all but the lowest redshifts. 

The high sensitivity of large single dish radio telescopes is an
advantage to those interested in studying
large scale baryon acoustic oscillations: \citet{Chang2010} report the
detection of \hi\ emission at a redshift of $z=0.8$ by cross
correlating the weak radio signal from the Robert C. Byrd Green Bank
Telescope with a map of the large scale structure known from optical
spectroscopic observations. 

Interferometric observations, with higher spatial resolution, are
able to resolve individual galaxies within large-scale structures. Utilising available
bandwidths, observations have successfully targeted narrow redshift
ranges around known galaxy clusters. In $\sim$2300 hours of observations
with the Westerbork Synthesis Radio Telescope (WSRT), the Blind
Ultra-Deep \hi\ Environmental Survey has detected more than 150 galaxies in
Abell 963 and Abell 2192 at $z=0.206$ and $z=0.188$, respectively
\citep{Jaffe2013}. 

Increased computing power and improvements in receiver technology are
advancing the field of radio astronomy by enabling
larger volumes to be surveyed to greater depth in vastly reduced time. 
This progress enables radio surveys to become competitive with
observations at other wavelengths in terms of both survey area and
accessible redshift.

Recent upgrades to the Karl G. Jansky Very Large Array (VLA) mean
is now possible to observe \hi\ over a continuous redshift range from
$0<z<0.45$. The CHILES (\textit{COSMOS \hi\ Large Extragalactic
  Survey}) collaboration has demonstrated the feasibility of such a
survey in a pilot project spanning $0<z<0.193$ \citep{Fernandez2013}.
APERture Tile in Focus (APERTIF), a focal-plane array system, will
increase the field of view of the WSRT, allowing faster mapping of
large areas of sky while measuring the \hi\ content of galaxies out to
$z=0.3$ \citep{Verheijen2008}. 

The upgrades to existing facilities serve as critical testbeds for
future facilities such as South Africa's MeerKAT (\citealt{Jonas2009})
and the Australian Square Kilometre Array Pathfinder (ASKAP,
\citealt{Johnston2008}), which themselves are to be 
precursor instruments for the international Square Kilometre Array
(SKA) Telescope. The SKA and the precursors will enable
observations of neutral hydrogen over cosmologically significant
ranges of redshift and to lower column densities than are possible
with existing instrumentation. 

Several surveys, such as Looking at the Distant Universe with the
MeerKAT Array (LADUMA, \citealt{Holwerda2012}), Deep Investigation of
Neutral Gas Origins
(DINGO\footnote{http://www.physics.uwa.edu.au/~mmeyer/dingo}),
Widefield ASKAP L-band Legacy All-sky Blind surveY
(WALLABY\footnote{http://www.atnf.csiro.au/research/WALLABY}) and 
deep APERTIF WSRT surveys are in advanced
planning stages and are set to begin within the next few years. These
surveys, requiring many thousands of hours of observing time, aim to
explore new parameter space and will operate at the limits of the
capabilities of the survey telescopes. However, even with the unprecedented sensitivity of
these instruments, direct \hi\ detections at moderate
($z\sim0.5$) redshifts will be rare due to the intrinsic weakness of
the emission. 

In order to extend the range of measurements to 
lower \hi\ column densities and higher redshifts, while keeping
observation times practical, techniques for
extracting information from datasets without statistical detections,
such as spectral stacking, are being investigated. These techniques
will be critical for the success of the above mentioned surveys,
particularly the projects aiming to study the neutral hydrogen content
of galaxies at $z\sim1$.

\subsection{\hi\ Spectral Stacking}

Stacking a large number of non-detections in order to build up a
single detection and recover the statistical properties of the
contributing ensemble of objects has long been used, primarily in
imaging data (\citealt{Dunne2009}, \citealt{Karim2011}). The spectral
dimension of a radio cube allows us to stack not only images but spectra as well. 
For the current work, we are only interested in detecting \hi\ 21cm line
emission from neutral hydrogen, not continuum emission resulting from
star formation or active galactic nucleus (AGN) activity. 

Radio spectral stacking relies on information from supplementary
observations, usually optical imaging and spectroscopy, to provide 
the sky positions and redshifts of a collection of galaxies. The
corresponding radio data-cube also contains the \hi\ emission from the
known galaxies, even if they are formally undetected. Spectra are extracted
from the cube at the locations of the known galaxies, shifted to the
rest-frame of each galaxy, and co-added, or stacked, to build up an
average detection from many non-detections. Section 3 in
\citet{Fabello2011} contains a comprehensive description of the 
stacking procedure.

In order to reach lower \hi\ mass sensitivity in less observing time,
\citet{Chengalur2001} undertook a deliberately shallow observation of
Abell 3128 ($z=0.06$) to test \hi\ spectral line stacking. This
technique has also been successfully employed by a number of 
groups to extend \hi\ studies to higher redshifts (\citealt{Lah2007}, 
\citealt{Lah2009}) and lower \hi\ content
(\citealt{Fabello2011}, \citealt{Delhaize2013}). \citet{Khandai2011} employ \textit{N}-body
simulations incorporating \hi\ to investigate the putative stacked signal at
$z\sim 1$ with encouraging results.

The primary assumption behind using optical redshifts for \hi\
spectral stacking is that the redshift determined from the optical
spectra is the same as that which would be determined from the \hi\
data, if it could be detected. This point is either assumed to be
approximately true with any differences averaging to zero
\citep{Fabello2011}, or the differences are sufficiently small
that they can largely be ignored \citep{Lah2007}. For galaxies
undergoing major mergers, or galaxies in overdense environments, the
stars and gas can have very different kinematic profiles
(\citealt{Haynes2007}, \citealt{Chung2009}), and the assumption of
identical optical and \hi\ redshifts will break down. 

The success of stacking also depends on the quality of the optical (or other
supplementary) astrometry and spectroscopy in particular. Accurate
redshifts of the galaxies are required in order to properly align the
non-detected galaxies in the spectral dimension. If the errors on the
measured redshifts are large, the \hi\ profiles will no longer be
aligned and the stacked profile will be smeared out.

The stacked signal is also sensitive to the noise properties of the radio
data. For a signal buried in Gaussian noise, stacking will increase
the flux in the signal linearly with the number of spectra
contributing to the stack ($n_{spec}$), but the noise increases only as the square
root of the number of spectra, thus the signal-to-noise ratio (S/N)
increases as $\sqrt{n_{spec}}$. This is not the case for non-Gaussian
noise.

The purpose of this work is to determine how closely redshifts derived
from optical spectroscopy relate to those measured from \hi\ profiles
using a sample of galaxies that are detected at high significance in
both optical and radio data. We are particularly interested in
determining the width of the distribution of the redshift
differences, as well as any systematic offset that may be present. 
We compare a number of methods of measuring the optical redshift to
assess whether particular measurements have advantages over others. 
Finally, we employ simple simulations to determine the effect this
redshift difference has on a putative stacked profile, and finish with
recommendations of minimum data quality requirements for future
stacking endeavours.

The outline of the paper is as follows. Section~\ref{sec:data}
describes the data used for the study, and Section~\ref{sec:redshifts}
compares the redshifts derived from optical and radio observations of
the same galaxies. Section~\ref{sec:stacking} introduces the simple
stacking simulation we employ to investigate redshift offsets between the
optical and radio observations, Section~\ref{sec:results}
describes the results of our simulations and
Section~\ref{sec:discussion} puts the results in the context of
upcoming surveys. Our conclusions are given in
Section~\ref{sec:conclusions}.

\section{Input \hi\ and Optical Data}\label{sec:data}

The \hi\ data come from the ALFALFA
survey (\citealt{Giovanelli2005a}, \citealt{Giovanelli2005b}, and
\citealt{Giovanelli2007}), specifically the $\alpha$.40 \hi\ source
catalog from \citet{Haynes2011}. The parameters of most interest for
the present study are the measured \hi\ recession velocity and the
velocity width of the \hi\ line profile measured at the 50~per~cent 
level of the peak, $W_{50}$.

The optical photometry and spectroscopy are from the Sloan Digital Sky
Survey (SDSS, \citealt{York2000}) Data Release 7 (DR7,
\citealt{Abazajian2009}).
The ALFALFA team has carefully crossmatched the $\alpha$.40 \hi\ detections 
to the SDSS database to determine the optical
counterpart for each. We include only the SDSS matches that also have
an optical spectrum, resulting in 9974 matches. From this crossmatch,
the SDSS ObjID and SpecObjID identifiers are known, and further
information from the SDSS database can be extracted. 

Flags from the crossmatch indicate a secure counterpart, or some 
confusion regarding the photometric or spectroscopic data. We only use
galaxies with an unambiguous optical counterpart possessing an SDSS
spectrum from the morphological centre of the galaxy to create a clean
starting sample, reducing the sample size to 9578. 

The sample is further reduced by imposing \vsys$\le15\,000$\,\kms\ to cut
out \hi\ data heavily affected by radio frequency interference
(RFI, \citealt{Martin2010}). Additionally, galaxies with an active 
nucleus and broad emission lines,
identified with SDSS spectral classification `AGN', are removed,
leaving 8923 galaxies.

The quality of the \hi\ detection determined by visual inspection of
individual \hi\ spectra is set as a flag within the ALFALFA
catalogue. Code 1 implies a reliable detection of high S/N, whereas
Code 2 detections generally have lower S/N but a plausible optical counterpart 
lending support to the detection. Imposing a straight S/N cut results
in a similar, but not identical, separation of the two categories.

The Code 1--2 designation is based on the quality of the \hi\ spectra,
and is not a quantity intrinsic to the \hi\ properties. We therefore
exclude the Code 2 ALFALFA galaxies from the sample, as
we aim to investigate the offset between optical and \hi\ velocities,
independent of problems arising from their measurement. This final
restriction leaves 6419 galaxies, and we refer to this subset of
ALFALFA galaxies with secure optical counterparts and high confidence \hi\
parameters as the ALFALFA--SDSS sample.

\section{\hi\ vs Optical Redshifts}\label{sec:redshifts}

We wish to compare the redshifts, or equivalently, the recession
velocities, of the ALFALFA--SDSS galaxies as determined from the \hi\
and the optical spectra. First, we investigate  
each of the optical and \hi\ measures separately, to gain confidence in
their reliability. Throughout, we convert between redshift and
recession velocity with $v=cz$, where $c=299792.458$\,\kms, as do
\citet{Haynes2011} for the \hi\ velocity measurements. Both the SDSS
and the ALFALFA velocities are set to the heliocentric reference frame.

\subsection{Spectroscopic Reliability}

HIPASS (\citealt{Barnes2001},
\citealt{Meyer2004}), the Equatorial Survey (ES, \citealt{Garcia2009})
subset, and the Northern HIPASS catalogue (NHICAT,
\citealt{Wong2006}) serve as an independent check of the reliability
of the recession velocities derived for the ALFALFA galaxies. There
are 142 unambiguous matches between HIPASS ES and NHICAT and
ALFALFA, as determined from inspection of the SDSS imaging, of which
109 are matches with ALFALFA optical counterpart flag `I', indicating
a single optical counterpart has been identified. The HIPASS
observations are not as sensitive as ALFALFA, 
so only the most \hi-rich galaxies appear in the overlapping sample,
and the S/N of the ALFALFA \hi\ detections is high. The spectral
resolution of HIPASS, at 18\,\kms, is also lower than the 11\,\kms\
resolution of ALFALFA. 

\begin{figure}
\resizebox{\hsize}{!}{\includegraphics{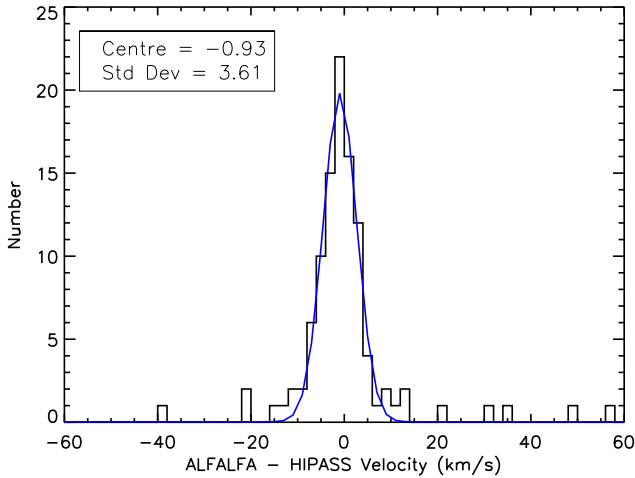}}
\caption{The measured recession velocity differences for ALFALFA
  galaxies matched to HIPASS ES and NHICAT to determine the
  repeatability of the ALFALFA redshifts. The centre and the standard
  deviation of the Guassian fit to the distribution are given in the plot legend.}
\label{fig:hi_comparison}
\end{figure}

As seen in Fig.~\ref{fig:hi_comparison}, the measured systemic velocities of
galaxies common to both ALFALFA and HIPASS correspond very well. There
are a few large velocity difference outliers, and visual inspection of
the SDSS images shows that there is always another galaxy within the
larger HIPASS beam that could be causing confusion.

\begin{figure}
\resizebox{\hsize}{!}{\includegraphics{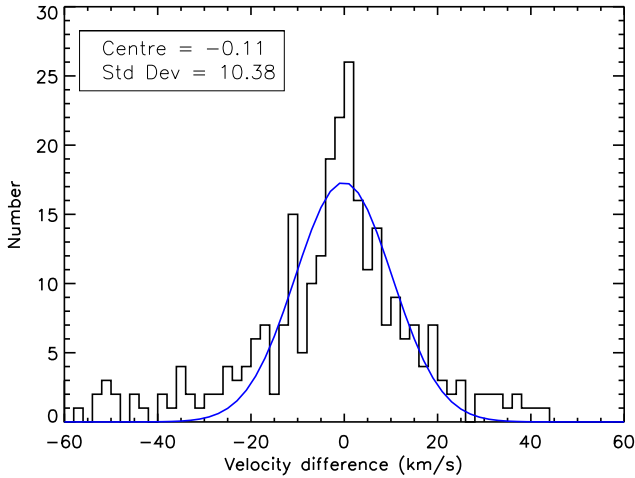}}
\caption{The velocity difference for the 240 galaxies in ALFALFA with
  more than one SDSS spectrum at the same position on the sky
  (i.e. repeat observations). The centre and the standard deviation of the Guassian fit
to the distribution are given in the plot legend.} 
\label{fig:doubleSDSS}
\end{figure}

A similar exercise can be done for the optical redshifts, as a number of
galaxies have repeat observations, i.e. two or more spectra at the
same sky position observed on different nights. There are 240 ALFALFA
galaxies with repeat observations, with the difference in recession
velocities derived from each pair of spectra shown in
Fig.~\ref{fig:doubleSDSS}. The width of this distribution,
$\sim$10\,\kms, is much smaller than the 60\,\kms\ error quoted by the
SDSS data releases, possibly because the ALFALFA--SDSS
galaxies of interest here are all at $z\le0.05$ and are generally
bright. Therefore, we conclude that, at least for the \hi-rich and
optically bright galaxies, we can treat both the \hi\ and the optical
recession velocities as highly reproducable.

\subsection{Optical Redshift Measurements}

The SDSS provides several redshift measurements for each spectrum
derived via independent methods, stored in different tables in the
Catalog Archive Server (CAS) database. Since we are interested in
the differences between radio and optical redshift determinations, we
need to ensure that we are using the best measurements available.

For each spectrum, the redshifts derived from fitting the emission
lines are stored in the 
ELRedshift table, with an associated confidence. The SpecLine table
stores the fit parameters to the individual measured emission 
lines used to derive the ELRedshift result. Single Gaussians are fit
to individual features, which become inappropriate for non-Gaussian
shaped features.

\begin{figure}
\resizebox{\hsize}{!}{\includegraphics{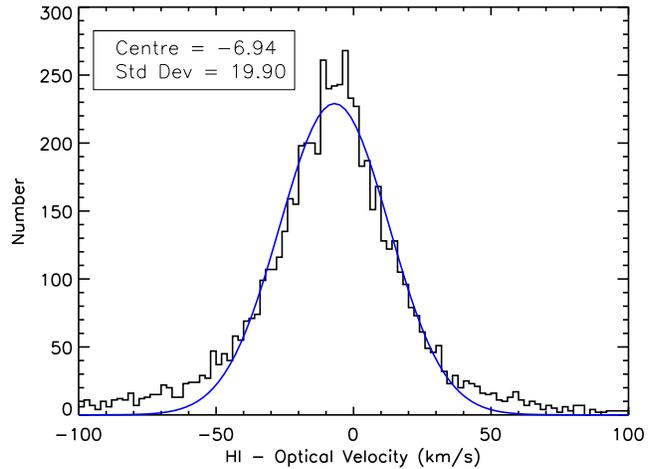}}
\caption{The difference between the velocity derived from the \hi\
  and that from the SDSS SpecObj $z$, based on the ALFALFA--SDSS
  galaxies. The centre and the standard 
  deviation of the Gaussian fit are given in the plot legend.}
\label{fig:HI-CAS}
\end{figure}

Alternatively, a cross-correlation of spectral templates to the spectra is also
performed, masking out emission features and using only absorption
features. The resulting redshift and its error is stored in the
XCRedshift table. The redshift from the SpecObj table, simply listed
with the variable name $z$, is the redshift determined from the emission lines or
cross-correlation method, whichever has the higher confidence. Fig.~\ref{fig:HI-CAS}
compares the recession velocity from the SpecObj $z$ and that from the
\hi\ measurements for the ALFALFA--SDSS galaxies. The distribution is
approximately Gaussian, and the resulting fit is offset from zero by
-6.9\,\kms\ and has width $\sigma=19.9$\,\kms. This distribution is
similar to that found by \citet{Toribio2011}, who also compared
optical and \hi\ recession velocities for ALFALFA galaxies with SDSS 
spectroscopy, and found a distribution with dispersion
$\sim35$\,\kms. The reduced width found here is due to
our exclusion of Code 2 ALFALFA detections and requirement that the
SDSS spectrum be centred on the galaxy. There 
are non-Gaussian wings of the distribution in Fig.~\ref{fig:HI-CAS},
with non-negligible numbers of galaxies having velocity differences
$>|40|$\,\kms. These are discussed further in Section~\ref{subsec:outliers}. 

There is a known velocity offset of 7.3\,\kms\ of the reported SpecObj $z$
redshifts from the SDSS, in the sense that the stated values are too low \citep{Adelman2008}.
The origin of the offset is still unknown, but it persists
in the DR7 data. If a correction is applied to the redshifts in
Fig.~\ref{fig:HI-CAS}, the offset from zero becomes -14.2\,\kms.
This offset is corrected in the sppParams table,
which uses the Spectro Parameter Pipeline (spp) processing instead of
the spectro1d pipeline which results in the SpecObj $z$ value. 

Lastly, a collaboration of researchers at the Max Planck Institute
for Astrophysics (MPA) and the Johns Hopkins University (JHU) have produced
the MPA-JHU value-added galaxy catalogue
(hereinafter referred to as the \textit{MPA--JHU catalogue}), which
provides spectral properties derived from an independent analysis
of the SDSS DR7 galaxy spectra. Within the MPA--JHU catalogue, a stellar
population model is fit to the galaxy continuum to properly account
for absorption features, which can be significant in some
cases. Further details of the fitting procedure can be found in
\citet{Tremonti2004} and \citet{Brinchmann2004}, or at the website hosting the
catalogues\footnote{http://www.strw.leidenuniv.nl/~jarle/SDSS}. The
redshifts from the sppParams table agree very well with those from 
the MPA-JHU catalogue. 

\begin{figure}
\resizebox{\hsize}{!}{\includegraphics{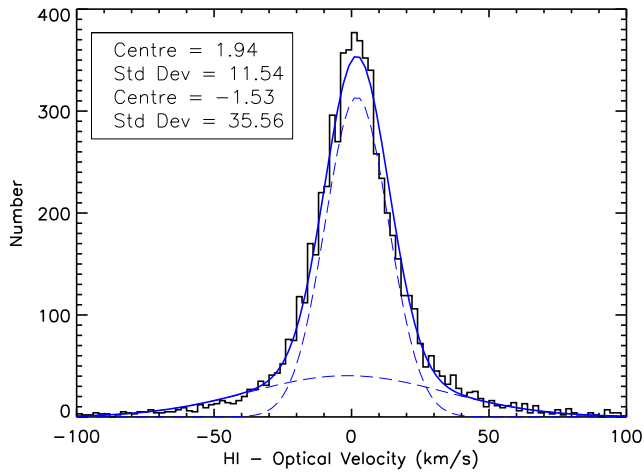}}
\caption{The difference between the velocity derived from the \hi\ and
  that from the JHU--MPA reprocessed data. The centres and standard
  deviations of the two Gaussian fits (blue dashed lines) are given in
  the legend. The blue solid line is the sum of the two individual curves.}
\label{fig:HI-MPA2_code1}
\end{figure}

The comparison of MPA--JHU catalogue and \hi\
velocities is shown in Fig.~\ref{fig:HI-MPA2_code1}. The distribution is
much narrower than that in Fig.~\ref{fig:HI-CAS}, and the offset from
zero is greatly reduced. Fitting with two Gaussians
produces a very good fit to both the core and the wings of the
distribution. The two Gaussians may imply two distinct
populations contributing to the total sample, and it allows for a
simple and useful parametrization of the \hi-optical velocity
offsets. As the MPA--JHU catalogue does not suffer from the 
7.3\,\kms\ offset, and there are additional useful measurements
within the MPA--JHU catalogue, we use these values for the
optical redshifts of the ALFALFA--SDSS galaxies. 

\subsection{Velocity Outliers}\label{subsec:outliers}

Fig.~\ref{fig:HI-MPA2_code1} shows that the optical and the \hi\ velocities do not
match exactly, and there is a significant population of galaxies with
velocity offsets $>|40|$\,\kms, which marks where the
distribution is entirely composed of objects in the broader
distribution. We wish to know if there is any
observable property that can be exploited to reduce the width of the
distribution or reduce the number of velocity outliers. 

Eight~per~cent of the ALFALFA--SDSS galaxies have \hi-optical
velocity differences $>|40|$\,\kms.
We investigated several parameters in an attempt to isolate
the outliers, including the width of the \hi\ profile (measured as
$W_{50}$), \hi\ flux, \hi\ S/N, \hi\ mass and the errors associated
with each of these measures,
the optical $u-r$ colour, optical galaxy orientation, and optical spectrum
S/N. While there is tentative evidence that the velocity outliers have
widths $W_{50}$ greater than the median width of 150\,\kms, from
visual inspection of the \hi\ and optical spectra, there is no
obvious trend for any of these parameters to preferentially select
velocity outliers.

Nearly ten~per~cent of the ALFALFA galaxies have publicly available \hi\
spectra accessible
online\footnote{http://arecibo.tc.cornell.edu/hiarchive/alfalfa/},
and more will become available as the \hi\ archive is updated.
Visual inspection of the outlier \hi\ profiles combined with the optical images
from SDSS and DSS2 can help determine the causes of the large offset
between the \hi\ and optical velocities. We find a number of factors which may be
responsible for large \hi--optical offsets:

\begin{itemize}

\item{Asymmetric \hi\ profiles -- The majority of cases fall within this
  category. The intrinsic \hi\ profile may be
  severely aysmmetric, leaving one side undetected.  Alternatively, the
  emission from the systemic velocity of the galaxy may only be at 
  the same level as the noise, making it difficult to recognize two
  peaks in the spectrum belonging to the same \hi\
  profile.}

\item{Clearly disturbed, interacting systems -- High S/N 
  \hi\ emission from a tidal tail or infalling gas
  contributes to the width of the \hi\ profile.  These are likely
  outliers because the systemic velocity of the galaxy is determined
  by the midpoint of the $W_{50}$ profile, whereas the optical redshift is
  determined from a fibre on the morphological centre of the
  galaxy.}

\item{Blended \hi\ profiles -- More than one galaxy in the Arecibo
    beam, with \hi\ profiles overlapping in velocity space, but the
    galaxies are not necessarily interacting. The
    optical spectrum will generally only measure the recession
    velocity for one of the galaxies, whereas the \hi\ profile
    may have contributions from several components.}

\item{Mis-identified \hi\ profiles -- These are extremely rare among
  the Code 1 sources, and may be due to difficult baseline
  subtraction, a portion of the data having lower weight, or the
  presence of RFI.}

\item{Absorption line spectrum -- A number of galaxies have significant
  reservoirs of \hi\ but do not show emission lines in their optical
  spectrum.  We will discuss in Section~\ref{subsec:lines} that optical
  redshift measurements derived from absorption features are more
  uncertain.}

\end{itemize}

In the context of \hi\ stacking, only the interacting systems
could result in an intrinsic mismatch between the optical and \hi\
measured systemic velocities. The fraction of interacting
systems will increase at higher redshifts
(\citealt{Ravel2009}, \citealt{Conselice2009}), so the number of objects with large
velocity offsets will increase with increasing redshift. The other effects result from issues
related to measuring quantities from the data. The absorption line
spectrum systems must also be treated with caution, as described in
Section~\ref{subsec:lines}. Asymmetric \hi\ profiles will not be
as much of an issue, since the \hi\ profile will be undetected, and the redshift
estimate will come from the optical spectrum alone. 

\subsection{Individual Spectral Features}\label{subsec:lines}

With future \hi\ stacking efforts in mind, the useful observable
properties must come from the optical photometry used for positional
information, or spectroscopic observations for the redshifts, as by
definition, the \hi\ will be undetected. Here, we 
investigate the spectral properties of the galaxies to determine
whether particular spectral features correlate more closely with the
\hi\ velocity.

\begin{figure*}
\resizebox{0.9\hsize}{!}{\includegraphics{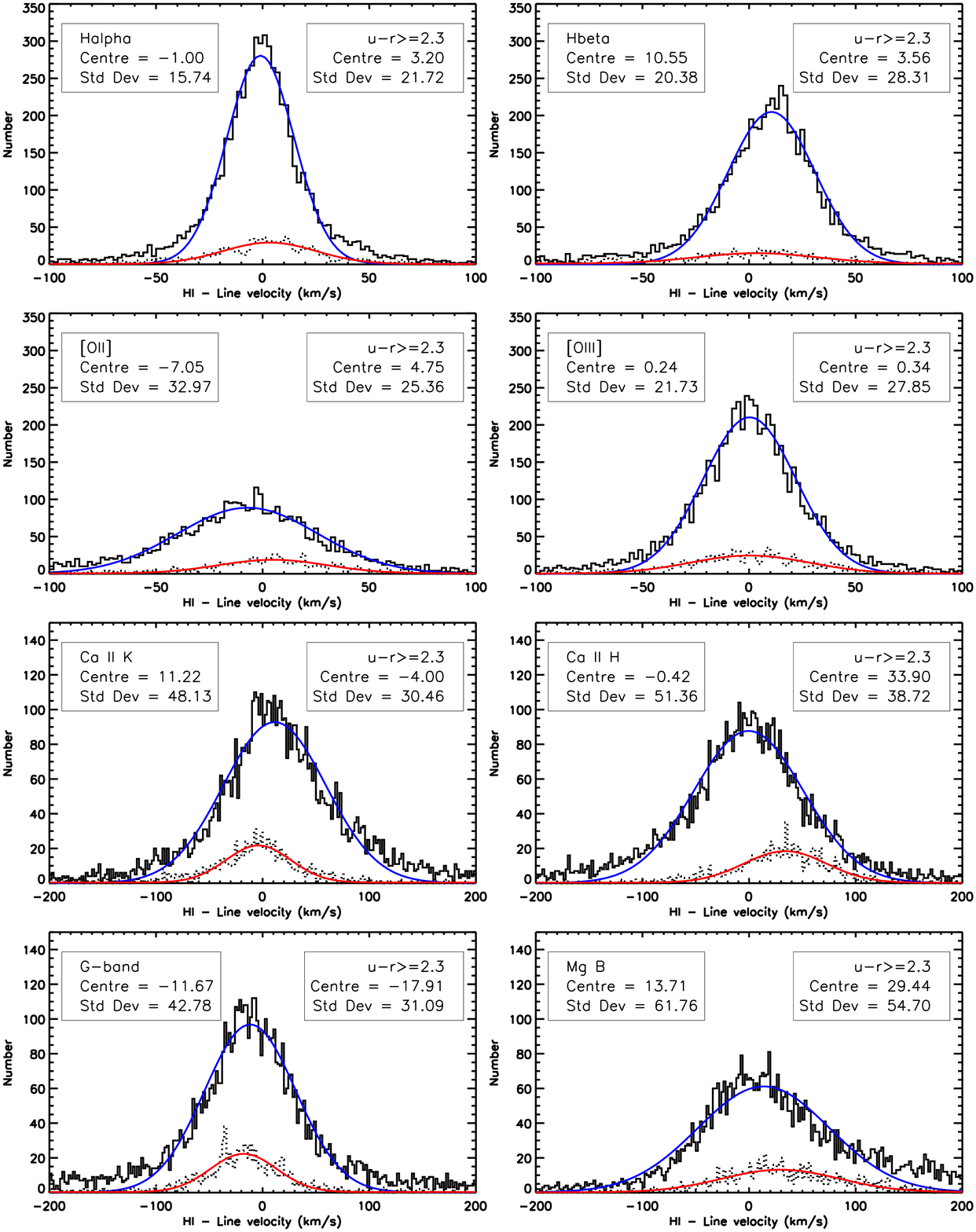}}
\caption{Comparison of velocities derived from individual spectral
  lines, compared to the ALFALFA \hi\ velocity. The spectral feature is
  listed in the left-hand caption, along with the centre and width of
  the distributions. The emission lines (top four panels) all
  have significantly narrower distributions than the absorption
  lines (bottom four panels). The sub-samples in each panel show the reddest galaxies with
$u-r\ge 2.3$, with the centres and widths of these distributions
  listed in the right-hand captions. The blue and red lines show
  the Gaussian curves fit to the full and red subsamples,
  respectively. Note the different x- and y-axis 
  ranges for the emission and absorption panels.}
\label{fig:HI-lines}
\end{figure*}

Fig.~\ref{fig:HI-lines} shows the difference between the \hi\
systemic velocity and the \vsys\ measured from individual spectral features,
extracted from the SDSS SpecLine table. Ideally, the comparison would
come from the MPA--JHU catalogue, but measurements for all the individual
lines are not available. Four emission lines and four
absorption features are shown. The emission lines in general result in
narrower distributions than the absorption lines, with the exception
of \oii\ 3727\AA, which for $z>0.02$ appears in the very blue 
end of the SDSS DR7 spectra with a blue wavelength limit of 3800\AA, 
and for $z\le 0.02$ is not covered by the SDSS spectra at all. The
distibution centre and standard deviation from a single Gaussian fit
to the histograms for each spectral feature in
Fig.~\ref{fig:HI-lines} is tabulated in Table \ref{tab:centres}.

Optical velocities determined from measuring the \halpha\ emission line
results in the narrowest distribution with a small offset. The
distribution for \hbeta\ is wider, with the centre significantly offset
from zero. This may be the result of the emission line sitting in a
corresponding absorption trough complicating the fit. The absorption
lines result in broader distributions than the emission lines.

The small histograms plotted in each panel are the subset of objects
that are red, with $u-r\ge 2.3$. This colour cut approximately divides the red
sequence galaxies from the blue cloud, as found in \citet{Baldry2004}
for a large sample of low redshift SDSS galaxies. The galaxies with $u-r\ge
2.3$ tend to be massive spirals with prominent bulges, dust lanes, and
viewed at high inclination angles. 

The centres of the velocity difference distributions for the full
sample and for the $u-r\ge 2.3$ subsample are different, sometimes by
a large amount, as for the Ca{\sc ii} H absorption feature. The
absorption lines of the red subsample of galaxies are wide with 
respect to the full sample of galaxies, while the widths of the
emission lines for the two subsamples show no such trend. This
indicates that the red galaxies have significant evolved stellar populations.
The velocity offsets are also much more pronounced for the absorption lines, 
further implying that the underlying cause must involve the older
stellar populations. 

The red galaxies have high S/N spectra with well-defined absorption
lines, so the distribution offsets are not due to poor measurement
fo the features. Measurement uncertainty would also result in wider
velocity difference distributions, not coherent shifts of the
distribution centres. Asymmetric or blended absorption line profiles could result
in systematic offsets of the resulting redshift measurement, if the
profile is fit with a single Gaussian, as is done within the SDSS
pipeline. Absorption features with either blue or red asymmetry can
result in positive or negative distribution offsets.

From Fig.~\ref{fig:HI-lines}, we conclude that measuring
redshifts from just one spectral feature is not as reliable as using
several features. As future \hi\ surveys reach higher
redshifts, spectral features will move through the observing window. 
H$\alpha$, which shows the narrowest distribution of all the lines,
will only be visible in the optical to $z<0.4$. \oiii is visible to
$z<0.8$, and for higher redshifts, \oii will be the only observable
prominent emission line, and only for galaxies with significant star
formation.

\begin{table}
\centering
\caption{Summary of the distribution centres and standard deviations
  for each of the spectral features shown in Fig.~\ref{fig:HI-lines}.}
\label{tab:centres}
\begin{tabular}{lcccc}\\ \hline
 &  &  & \multicolumn{2}{c}{$u-r\ge2.3$} \\
Spectral & Centre & Std Dev & Centre & Std Dev \\
Feature & \kms & \kms & \kms & \kms \\ \hline
H$\alpha$ & -1.00 & 15.74 & 3.20 & 21.72 \\
H$\beta$ & 10.55 & 20.38 & 3.56 & 28.31 \\
\oii 3727\,\AA & -7.05 & 32.97 & 4.75 & 25.36 \\
\oiii 5008\,\AA & 0.24 & 21.73 & 0.34 & 27.85 \\
Ca{\sc ii} K 3935\,\AA & 11.22 & 48.13 & -4.00 & 30.46 \\
Ca{\sc ii} H 3970\,\AA & -0.42 & 51.36 & 33.90 & 38.72 \\
G-band 4306\,\AA & -11.67 & 42.78 & -17.91 & 31.09 \\
Mg B 5177\,\AA & 13.71 & 61.76 & 29.44 & 54.70 \\ \hline
\end{tabular}
\end{table}

\section{Simple Stacking Simulation}\label{sec:stacking}

In order to quantify the effect the velocity difference distribution
shown in Fig.~\ref{fig:HI-MPA2_code1} has
on a putative stacked \hi\ signal, we have constructed a simple spectral co-adding
simulation. The simulation does not account for issues arising from
the \hi\ observations, such as RFI. We are purely interested in assessing the relative
properties of the resulting stacked \hi\ profiles when issues related
to the accuracy of the optical spectrosocpy are incorporated.

The inputs of the simulation are the noise properties of
the \hi\ spectrum of a non-detection, a distribution of velocity
offsets to represent the difference between the \hi\ and optically
determined redshifts, and a distribution of intrinsic \hi\ widths. The
noise is set to be Gaussian throughout, although there are indications
that at some level the noise of real observations is not purely
Gaussian (fig. 5 of \citealt{Fabello2011} is a good illustration of this).
The noise properties, in general, will be dependent on the instrument
and observing conditions. 

\begin{figure}
\resizebox{\hsize}{!}{\includegraphics{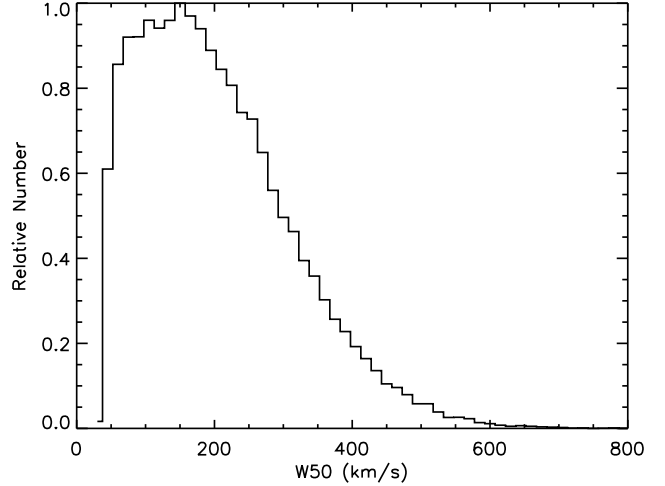}}
\caption{The input \hi\ $W_{50}$ width distribution for the stacking
  simulations, which is an approximation to the distribution shown in
  fig. 2b of \citet{Haynes2011}.}
\label{fig:sim_hiwidth}
\end{figure}

\begin{figure}
\resizebox{\hsize}{!}{\includegraphics{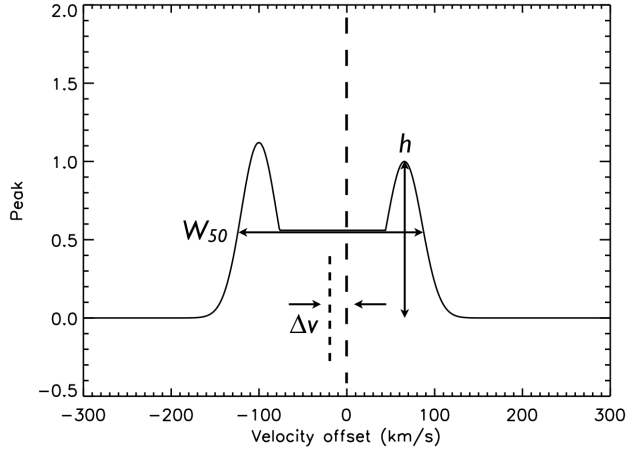}}
\caption{The variable parameters for the simulated \hi\ profiles for
  the stacking simulations. The height of one peak, \textit{h}, is set
to unity, with the height of the other peak varying with respect to
this. }
\label{fig:oneprofile}
\end{figure}

The \hi\ profiles are modelled as two Gaussian curves each of fixed width, with
separation either fixed at $W_{50}=150$\,\kms, or drawn from 
the distribution of $W_{50}$ shown in Fig.~\ref{fig:sim_hiwidth}, 
which closely approximates the \hi\ $W_{50}$ distribution from the
ALFALFA survey (see fig. 2b in \citealt{Haynes2011}). The two peaks of the \hi\ profile vary in
height with respect to each other, with about 30~per~cent of the profiles
having peaks differing in height by more than 20~per~cent, which is
broadly consistent the census of \hi\ profile asymmetry from
\citet{Richter1994}, who find 20~per~cent of profiles are strongly
asymmetric. A constant fills in the gap between the two profiles when it dips below half
of the peak height. An example \hi\ profile with the relevant features is shown in
Fig.~\ref{fig:oneprofile}.

The velocity offset is either set to zero to simulate the case
of identical optical and \hi\ redshifts, a combination of 11\,\kms\ and
36\,\kms\ based on Fig.~\ref{fig:HI-MPA2_code1}, or a series of increasing
values representative of possible optical observations. The largest
offset of 250\,\kms\ is loosely based on results from the 
redshift survey of the GOODS South field undertaken by
\citet{Balestra2010}. Using the VLT VIMOS 
instrument with the Low Resolution Blue grism (R$\sim$180),
\citet{Balestra2010} find the redshift consistency for galaxies with
more than one spectrum is measured to have a distribution of width
360\,\kms, with the accuracy on a single redshift measurement of
360/$\sqrt{2}$=255\,\kms. For the Medium Resolution orange grism 
(R$\sim$580), the redshift consistency distribution has width 168\,\kms,
and the accuracy for a single measurement of 120\,\kms. These determinations
of redshift accuracy indicate the difficulty of obtaining accurate
measures of high redshift galaxies, even with an 8-m class telescope.

We use two separate methods for setting the peak of the undetected \hi\
profiles with respect to the noise. In the first case, one
peak of the \hi\ profiles is given in terms of a fraction of the sigma
of the noise, while the other peak can be larger or smaller. Here, we
set the peak to be 1$\sigma$, to give a peak 
S/N $=1$. For convenience, we refer to these
profiles as the \textit{constant peak profiles}, and an example of
this profile type is shown in Fig.~\ref{fig:oneprofile}. 

The second case sets a reference profile of width $W_{50}=150$\,\kms\ and
peak height of 1$\sigma$. The integrated area under this curve is
computed and used as the reference area. We scale all the other curves of varying
widths such that the area under their curves is equal to the
reference area. Physically, as the area under the profile curve
is a measure of the \hi\ mass of a galaxy, this case describes an ensemble of
galaxies of equal \hi\ masses within a redshift shell. Narrower
profiles end up with peak S/N $>1$ and wider profiles have S/N $<1$, but
all have the same \textit{integrated} signal. We refer to these profiles
as the \textit{constant area profiles}. 

The constant peak and constant area profiles span a range of input
\hi\ mass distributions, from combining galaxies of different masses
(constant peak profiles) to combining galaxies all of the same mass
(constant area profiles), while maintaining the observationally
determined distribution of \hi\ profile widths. The results allow us to begin to understand
the relative importance of the input \hi\ mass distribution into the
stacked signal by comparing the resulting profiles. 

A number of simulations are presented here to facilitate
comparison. The parameters for each are listed in
Table~\ref{tab:sims}. Simulations with $W_{50}$ listed as Fixed are
unphysical as the profiles are all of identical widths, but serve as a
useful reference point and show the effect of the velocity offset
only. Similarly for simulations 2 and 3, the effect of the \hi\
distribution is isolated, as the velocity offset is set to zero. Simulations 5 and 6 represent the scenario
that most closely matches the data we have from the ALFALFA \hi\
profiles and the corresponding SDSS spectra. Simulations 20 and 21 show
the effect of much poorer quality optical spectra.

For each combination of parameters, we simulate 1000 spectra, with the
parameters drawn from the velocity offset and $W_{50}$
distributions. For the velocity offsets based on
Fig.~\ref{fig:HI-MPA2_code1}, parametrized with two Gaussians, one in
three offsets is chosen from the wide profile, to approximate the
relative contributions of the wide and narrow components.
The 1000 constructed spectra are stacked in no particular order, using
a simple averaging for each channel, following:

\begin{equation}
S = \frac{\sum_{i=1}^{1000}S_i w_i}{\sum_{i=1}^{1000}w_i}
\label{eq:stack}
\end{equation}

\noindent where $S_i$ are the individual spectra, $w_i$ are weights
for each spectrum as the inverse of the square of the spectrum RMS
noise, and $S$ is the stacked spectrum. For the
simulations, the weights are ignored as the noise of each spectrum is the same. The
progress of the resulting stacked spectrum is followed with each
additional spectrum. This process is repeated 50 times for each set of
parameters to ensure the distributions are fully sampled and the
results are averaged.

\begin{table}
\centering
\caption{Variables incorporated various runs of the
  simulations. Profile type = P indicates constant peak profiles, and
  Profile type = A indicates constant area profiles.}
\label{tab:sims}
\begin{tabular}{cccc}\\ \hline
Number & Velocity Offset & $W_{50}$  & Profile Type \\
 & Dist'n Width (\kms) & Dist'n & \\ \hline
1 & 0 & Fixed & P  \\
2 & 0 & Variable & P  \\
3 & 0 & Variable & A \\
4 & 11/36 & Fixed & P \\
5 & 11/36 & Variable & P \\
6 & 11/36 & Variable & A \\
7 & 60 & Fixed & P \\
8 & 60 & Variable & P \\
9 & 60 & Variable & A \\
10 & 100 & Fixed & P \\
11 & 100 & Variable & P \\
12 & 100 & Variable & A \\
13 & 150 & Fixed & P \\
14 & 150 & Variable & P \\
15 & 150 & Variable & A \\
16 & 200 & Fixed & P \\
17 & 200 & Variable & P \\
18 & 200 & Variable & A \\
19 & 250 & Fixed & P \\
20 & 250 & Variable & P \\
21 & 250 & Variable & A \\ \hline
\end{tabular}
\end{table}

\section{Simulated Stacked Spectra}\label{sec:results}

Here we describe the outcome of the simulations described in
Section~\ref{sec:stacking}, including the changes in profile shape,
how the stacked profile flux compares to the fluxes of the input
galaxies, and what we can learn about stacking in the context of
future survey design.

\subsection{The Stacked Profiles}

\begin{figure}
\resizebox{\hsize}{!}{\includegraphics{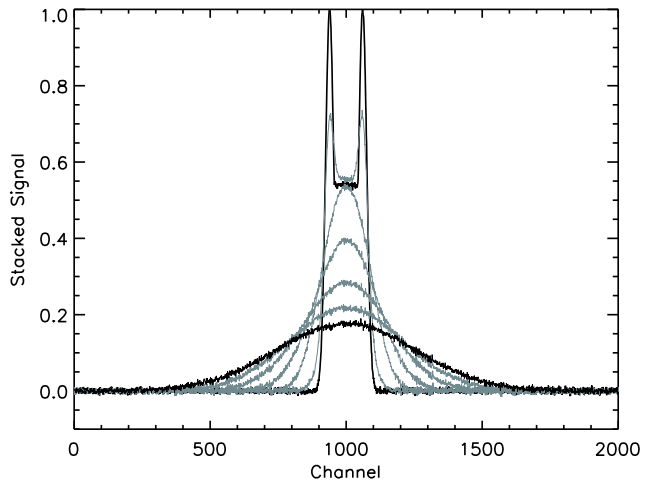}}
\caption{Comparison of the stacked profiles for 1000 constant peak
  profiles of fixed width ($W_{50}=150$\,\kms) and velocity offset
  distribution widths $\Delta v$=0, 11/36, 60, 100, 150, 200, 250
  \kms\ (top to bottom), averaged over 50 iterations. The x-axis is incremented by ones, so each
  point corresponds to a `channel' of width 1\,\kms. Each profile has
  been divided by 1000, the number of input spectra.}
\label{fig:profcompare0}
\end{figure}

\begin{figure}
\resizebox{\hsize}{!}{\includegraphics{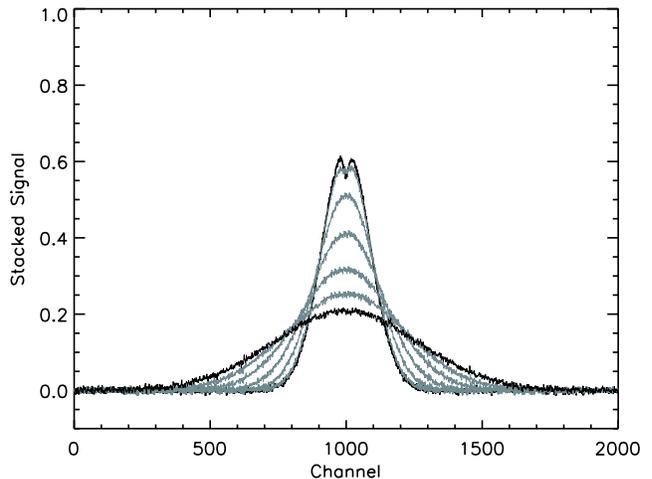}}
\caption{Comparison of the stacked profiles for 1000 constant peak
  profiles and velocity offsets $\Delta v$=0, 11/36, 60, 100, 150,
  200, 250 \kms\ (top to bottom), averaged over 50 iterations.}
\label{fig:profcompare}
\end{figure}

\begin{figure}
\resizebox{\hsize}{!}{\includegraphics{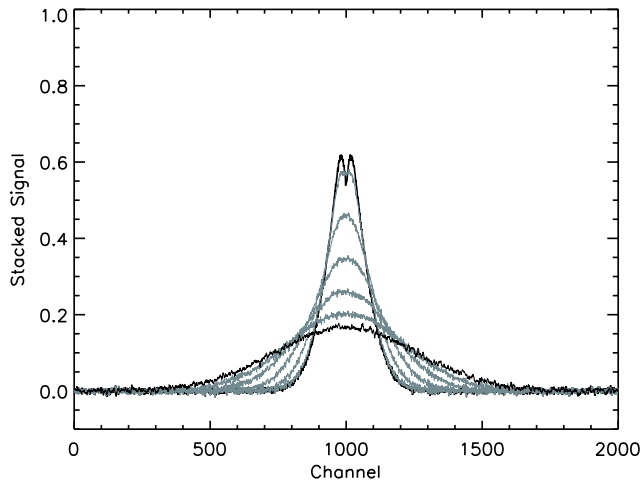}}
\caption{Comparison of the stacked profiles for 1000 constant area
  profiles and velocity offsets $\Delta v$=0, 11/36, 60, 100, 150,
  200, 250 \kms (top to bottom), averaged over 50 iterations.}
\label{fig:profcompare2}
\end{figure}

The stacked profiles from 1000 spectra corresponding to simulations 1,
4, 7, 10, 13, 16 and 19 (profiles of fixed width) are shown in
Fig.~\ref{fig:profcompare0}, divided by the number of input spectra,
and averaged over the 50 iterations. The
double-peaked profile shape is only visible while the velocity errors
are small. While these profiles are unrealistic as they are drawn from
a set of identical \hi\ profiles representing identical galaxies,
they isolate the effect of the redshift errors from the distribution of input profile
widths. Once the velocity errors are larger than 11\,\kms, the width of
the profile is no longer well measured by $W_{50}$, so we fit a
Gaussian to the curve and measure the standard
deviation. Table~\ref{tab:widths} lists the standard deviations for 
each of the profiles shown in Figs. ~\ref{fig:profcompare0} through \ref{fig:profcompare2}.
The noise measured in channels 0--300 and 1700--2000, away from the
signal for each stacked spectrum after combining $n$ spectra, retains Gaussian behaviour.

The stacked profiles from 1000 spectra corresponding to the constant
peak profiles with varying widths (simulations 2,
5, 8, 11, 14, 17 and 20) are shown in Fig.~\ref{fig:profcompare}. Note
that the input double-peaked profile is washed out in all of the
stacked profiles by the distribution of input profile widths, and the
width of the profiles with large $\Delta v$ become wider than the
median input width of $W_{50}=150$\,\kms. 

Fig.~\ref{fig:profcompare2} shows the stacked profiles
corresponding to the constant area profiles (simulations 3, 6, 9, 12,
15, 18 and 21). The
resulting \hi\ profiles are slightly narrower than for the constant peak
profiles, due to the relatively strong contribution from narrow input
profiles and the weaker contribution from wide profiles, but otherwise
the profiles in Figs.~\ref{fig:profcompare} and \ref{fig:profcompare2} are 
very similar. This indicates that the relative width of the stacked
profile is more sensitive to the input profile width distribution
rather than the input \hi\ mass distribution, particularly when the velocity
offset distribution width is large. 

\begin{table}
\centering
\caption{Standard deviations of the profiles shown in
  Figs. ~\ref{fig:profcompare0} through \ref{fig:profcompare2}. All
  values are in \kms.}
\label{tab:widths}
\begin{tabular}{cccc}\\ \hline
Velocity & Fixed & Constant & Constant \\ 
Offset & Width & Peak & Area \\ \hline
0 & 63 & 92 & 73 \\
11/36 & 68 & 95 & 76 \\
60 & 86 & 111 & 97 \\
100 & 117 & 138 & 127 \\
150 & 161 & 177 & 168 \\
200 & 209 & 222 & 216 \\
250 & 255 & 269 & 258 \\ \hline
\end{tabular}
\end{table}

\subsection{Derived Fluxes of Stacked Profiles}

One of the goals of \hi\ stacking is to derive an average mass for the
contributing galaxies. We therefore investigate how the measured mass
of the stacked profiles compares to the masses of
the input galaxies. As the relationship between flux and mass has a
straightforward distance dependence, we compute only fluxes for our
input and stacked profiles, with the understanding that the fluxes can easily
be converted into masses. With real data, redshifts to individual
contributing objects will be known from the optical spectroscopy.
 
The flux of each of the 1000 individual contributing profiles is measured before
adding noise by integrating over the 2000 channels. Both the average
and median of these input fluxes are calculated for comparison with
the flux in the stacked profile.

To compute the flux of the stacked profiles, we integrate under each
of the curves shown in Figs.~\ref{fig:profcompare0} through
\ref{fig:profcompare2}. For the narrow profiles with small velocity offsets,
the edges of the profiles are obvious, thus the limits of
integration are well determined. However, for the wider profiles, the
limits of the profile are not as well defined. Therefore, for each profile that is well fit
by a Gaussian, we find the centre and sigma of the fit, and integrate
over $\pm$3$\sigma$. For narrow, non-Gaussian shaped profiles, or profiles
with low S/N, we simply integrate over the entire
2000 channels, as regions without signal will average to zero since
the noise is Gaussian-distributed about zero. This may not be the case for
stacking real data, as low-level ripples from standing waves in single
dish data, such as that found by \citet{Fabello2011}, or low-level
RFI, sidelobes from undetected sources in the field, and other
calibration issues in interferometric data will cause the
stacked noise to be non-Gaussian.

We find that we are able to recover the average
flux of the input profiles from the stacked profile, with no
dependence on the velocity offsets, and thus the stacked profile width.
For the constant peak profiles with variable widths, the flux in the
stacked profile is larger than the average 
flux of the input profiles by 1--2 per~cent. It is larger than the
median flux of the input profiles by 6--7~per~cent, indicating that the
rarer, larger input profiles significantly contribute to the stacked
profile. The profiles with constant widths show similar
behaviour as the constant peak profiles. For
the constant area profiles, the average and median input 
profile fluxes are identical, and differ from the stacked profile flux
by less than one per~cent. Therefore, in these ideal cases, the flux
contained within the stacked profile is a good representation of the
average flux of the input profiles.

\subsection{Applications to Future Surveys}

The dependence of the stacked profile width on the quality of
optical redshifts is useful for justifying requirements for
spectroscopic ancillary data for upcoming \hi\ deep field surveys. We
have demonstrated that the width of a stacked \hi\ profile is 
dependent on the $W_{50}$ distribution of input \hi\ profiles for
small differences between the optical and \hi\ velocities. When the
velocity errors are comparable to or larger than the median $W_{50}$
width of the input profiles, they dominate the stacked profile width. 

By combining the results of the stacking simulations with a relationship between
the measured flux of a stacked signal and the integrated signal to
noise, we can predict the average \hi\ mass detectable from a stacked
profile at a particular redshift, given a specified integrated S/N 
and the parameters of the survey and instrumentation. 

We start with the equation for integrated S/N from \citet{Haynes2011}:

\begin{equation}
\frac{S}{N}=\frac{1000\, S_{21}}{W_{50}}\frac{\omega_{smo}^{1/2}}{\sigma}
\label{eq:intSNR}
\end{equation}

\noindent where $1000\, S_{21}$ is the integrated flux in mJy \kms, $W_{50}$ is
the full width at half max of the \hi\ profile, and $\sigma$ is the
noise in mJy of the signal-free portion of the spectrum. $\omega_{smo}$
is a `smoothing width', which for the ALFALFA survey is defined as
$\omega_{smo}=W_{50}/(2\times10)$ for profiles less than 400\,\kms\
wide, and 10 is the spectral resolution in \kms\ for the final ALFALFA cubes.
Hereinafter, we express the channel width as $\Delta w$ to generalize the smoothing
width to any survey.

Equation~\ref{eq:intSNR} is appropriate for individual \hi\ profiles with
relatively steep sides, where $W_{50}$ is a good measure of the width
of the profile and encompasses most of the flux. However, as seen in 
Figs.~\ref{fig:profcompare0}, \ref{fig:profcompare} and \ref{fig:profcompare2}, the sides of
stacked signals are not steep, and $W_{50}$ is no longer a good
approximation to the full width over which flux is measured. In order
to include the majority of the flux in the stacked profiles, 
instead of $W_{50}$, we use $W_{full}$, which we define as $\pm3$ times
the standard deviation of a Gaussian function fit to the stacked
\hi\ profile. This is also the range over which we integrate the
stacked profiles from the simulations to compute the total
flux. Equation~\ref{eq:intSNR} is then rewritten as: 

\begin{equation}
\frac{S}{N} = \frac{1000\, S_{21}}{\sigma\, W_{full}} \left(\frac{W_{full}}{2\, \Delta w}\right)^{1/2}.
\label{eq:intSNR_alt}
\end{equation}

\begin{figure}
\resizebox{\hsize}{!}{\includegraphics{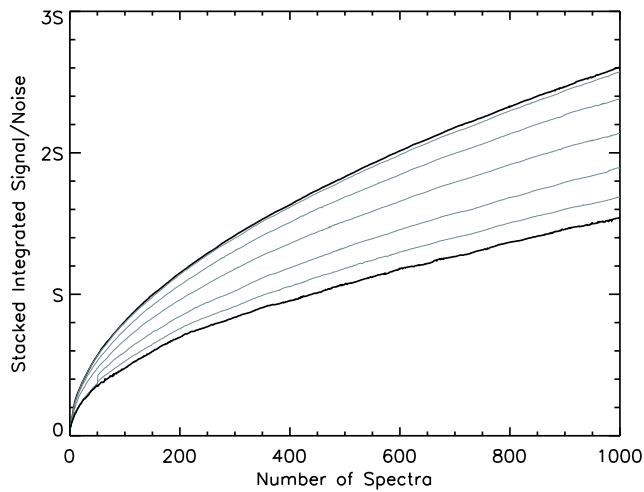}}
\caption{Comparison of the integrated S/N of the profiles shown in
  Fig.~\ref{fig:profcompare} for velocity offsets $\Delta
  v$=0, 11/36, 60, 100, 150, 200, 250 \kms\ (top to bottom). The integrated S/N value at
1000 spectra corresponds to the profiles shown in
Fig.~\ref{fig:profcompare}. The scale on the y-axis is arbitrary, with
units of S representing integrated S/N.}
\label{fig:snrcompare}
\end{figure}

A measure of the integrated S/N of the stacked \hi\ constant
peak profiles and how it grows with adding more stacked spectra is shown in
Fig.~\ref{fig:snrcompare}. The integrated S/N grows faster for
narrower profiles. The $S/N$ for each of the three profile types
show similar behaviour. For large velocity offsets and small numbers of
input spectra, the stacked profile is not well-defined, so we simply
integrate the stacked signal between 300--1700 channels. The
transition between integrating over a set range and integrating over
$W_{full}$ is visible for small numbers of input spectra as a
discontinuity in the curve. The noise
does not contribute significantly to the total flux.

For \hi\ masses at any redshift, the well-known \hi\ mass equation is:

\begin{equation}
\frac{M_{HI}}{M_{\odot}} = \frac{2.36\times10^5\, D^2}{1+z} \int S_{v} \,d v \\
\label{eq:HImass}
\end{equation}

\noindent where $M_{HI}$ is the total \hi\ mass in solar masses, $D$
is the luminosity distance to the object in Mpc, and the integral is
the total flux in Jy \kms, equivalent to $S_{21}$ in
equations~\ref{eq:intSNR} and \ref{eq:intSNR_alt}.  The $1+z$ factor
accounts for the difference between the observed and rest-frame width
of the measured profile. Rewriting equation~\ref{eq:HImass} in terms
of equation~\ref{eq:intSNR_alt}: 

\begin{eqnarray}
\frac{M_{HI}}{M_{\odot}} & = & \frac{2.36\times10^5\, D^2}{1+z} S_{21} \\
       & = & \frac{2.36\times10^5\, D^2}{1+z}\, \frac{(2\, \Delta w\, W_{full})^{1/2}\, \sigma}{1000}\, \frac{S}{N}.
\end{eqnarray}

This is the total \hi\ mass in solar masses for a given
profile. Assuming that the noise is fairly flat over the frequency
range surveyed, which is not unreasonable (\citealt{Fernandez2013}),
then the noise of the stacked spectrum is related to the noise of a
single spectrum (the inherent noise of a survey, $\sigma_1$) by
$\sigma=\sigma_1 n_{spec}^{1/2}$.  By dividing both sides of the mass
equation by the number of spectra stacked, we can express the average
\hi\ mass per contributing object from a stacked signal as:

\begin{equation}
\left<\frac{M_{HI}}{M_{\odot}}\right>=2.36\times10^5\,\left<\frac{ D^2}{1+z}\right> \frac{(2\, \Delta w\, W_{full})^{1/2}}{1000}\, \frac{\sigma_1}{n^{1/2}_{spec}}\, \frac{S}{N}.
\label{eq:masssensitivity}
\end{equation}

The advantage of expressing the \hi\ mass in this way is that all but
one of the variables are either defined by the
observation parameters ($\Delta w$, $\sigma_1$) or they can be
estimated for a specific scenario, for example the number of spectra
available to stack in a given redshift shell ($D$ and $1+z$, $S/N$,
$n_{spec}$). The remaining variable, $W_{full}$, can be estimated from
our stacking simulations. From Figs.~\ref{fig:profcompare0} through
\ref{fig:profcompare2}, $W_{full}$ depends on the distribution of optical--\hi\
velocity differences for the galaxies input into the stacked profile,
parametrized by the velocity offset in the stacking simulations.

\begin{figure}
\resizebox{\hsize}{!}{\includegraphics{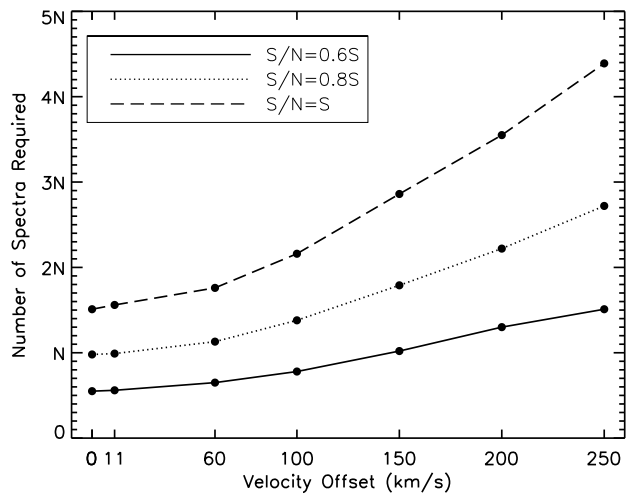}}
\caption{Comparison of the relative number of spectra required to achieve an
  integrated S/N=0.6S, 0.8S and S for velocity offsets $\Delta
  v$=0, 11/36, 60, 100, 150, 200, 250 \kms\ and the constant peak profiles.}
\label{fig:snrspec}
\end{figure}

As seen in Fig.~\ref{fig:snrcompare}, the number of spectra required
to reach a given integrated S/N increases with increasing stacked profile
width. Fig.~\ref{fig:snrspec} shows the relative number of constant
peak profile spectra required to
reach $S/N=$ 0.6S, 0.8S and S, which is equivalent to making three
horizontal cuts in Fig.~\ref{fig:snrcompare}. The behaviour is similar
for the constant width and constant area profiles. Approximately twice as many
spectra are required for the largest velocity offsets compared with
velocity offsets $<100$\,\kms. For large $S/N$, an arbitrarily large
number of spectra would be required for the widest
profiles. Conversely, given a finite number of spectra available for
stacking, the integrated S/N that can be achieved is reduced for larger velocity
offsets. 

\begin{figure}
\resizebox{\hsize}{!}{\includegraphics{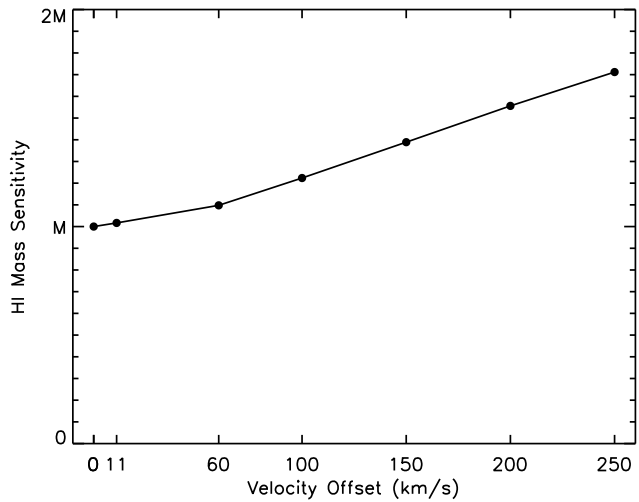}}
\caption{Comparison of the \hi\ mass sensitivity for 1000 spectra and
  integrated S/N=S, for velocity offsets $\Delta
  v$=0, 11/36, 60, 100, 150, 200, 250 \kms.}
\label{fig:mhisens}
\end{figure}

Fig.~\ref{fig:mhisens} shows how the relative \hi\ mass sensitivity
changes with increasing stacked \hi\ profile width for the constant
peak profiles. This is effectively taking a vertical cut through
Fig.~\ref{fig:snrcompare}. For a given redshift, the \hi\ mass 
sensitivity achieved with a given number of spectra improves by a factor
of $\sim$1.7 as the width of the \hi\ profile decreases.

\section{Discussion}\label{sec:discussion}

The results of the ALFALFA--SDSS redshift comparison and the
simulations enable us to make some general recommendations regarding
observation strategies in preparation for surveys that will employ
\hi\ stacking. As the \hi\ will be undetected, any useful quantities
must be derived from the optical data. We also discuss how the results
presented here will apply to future telescopes and surveys at high redshift.

\subsection{General Recommendations}

The success of \hi\ stacking depends heavily on the quality of
the optical spectroscopy. The spectral resolution of the SDSS spectra,
at about R=1800--2200, is sufficient to resolve spectral features such
as the \halpha, [N\,{\sc ii}] and \hbeta, \oiii emission line groups. 
From the ALFALFA--SDSS comparison, these spectra, with accurate
wavelength calibration, give redshifts that match the \hi\ redshifts
to within tens of \kms. Measuring a single spectral feature does not give
as accurate redshifts as using a number of features, but from the
simulations, we see that as long as the difference between the optical
and \hi\ redshifts are less than the median width of the \hi\
profiles, the stacked profile shape is largely unaffected by the
redshift uncertainties. This is in qualitative agreement with work by
\citet{Khandai2011}, who find that redshift errors of $\Delta z <
35$\,\kms\ dilute their simulated stacked profile peak by less than 3~per~cent.

Photometric redshifts are one technique used to circumvent 
telescope-intensive spectroscopy, and have successfully been applied
to extract a number of cosmological results (see, for example,
the COMBO-17 project, \citealt{Wolf2004}). However, even discarding the galaxies assigned
redshifts that greatly differ from the real value, referred to as
catastrophic outliers, the redshift accuracy achieved is of the order
$\Delta z/(1+z)> 0.01$, corresponding to velocity uncertainties
of several thousands of \kms. From Figs.~\ref{fig:profcompare} and
\ref{fig:profcompare2}, the resulting profile will be unreasonably wide,
and the integrated SNR will hardly increase with
additional stacked objects, using the simple stacking
procedure followed here. However, photometric redshifts will
certainly be useful for pre-selection of galaxy targets in redshift shells, allowing highly
efficient follow-up observations with moderate resolution spectroscopy.

Galaxies with red optical colours also seem to produce less accurate redshifts.
It may be tempting to target only blue,
star-forming galaxies and ignore the red, absorption line galaxies in
spectroscopic campaigns, but there is evidence that spheroids contain
a non-negligible amount of cold gas (see, for example,
\citealt{Grossi2009}, \citealt{Serra2012}), and should thus be included in an
\hi\ census. However, if the goal is to construct the cleanest stacked
spectrum possible, excluding red galaxies is a possibility for
particular cases.

As the current and future surveys will encompass large volumes,
acquiring spectroscopy of every galaxy in the survey volume is
unrealistic. It also may be large effort for low gain, as can be seen in
Fig.~\ref{fig:snrcompare}. For small numbers of spectra, the
integrated S/N increases rapidly, but after a few hundred spectra, the
S/N increases much slower with additional spectra. This is
particularly true for lower quality (large velocity error) spectra. Therefore,
consideration of the number of
galaxies available for stacking, and the quality of the optical
spectra to be collected, is necessary for planning the ancillary data
collection. Equation~\ref{eq:masssensitivity} will be useful for this planning.

\subsection{Future Telescopes and Surveys at High Redshift}

The results presented in this paper are based on galaxies in the local
Universe, observed with a single-dish radio telescope. Ongoing and upcoming
surveys are almost exclusively being undertaken with interferometers,
such as the CHILES project with the JVLA, LADUMA with MeerKAT, and
DINGO with ASKAP. These facilities all have significantly better spatial resolution
than Arecibo. This will be useful for separating close pairs of
galaxies which would be unresolved in single dish data.

The expanded frequency range of upcoming facilities, particularly of
MeerKAT, will enable \hi\ studies to progress to $z\sim1.4$. The galaxy 
population has undergone significant evolution from $z=1$ to the
present, so it is
interesting to consider how the results from this $z\sim0$ study may
apply at high redshifts. The fraction of interacting galaxies is known to
increase with redshift (\citealt{Conselice2009}), which will contribute to
the outlier population in the optical--\hi\ redshift differences,
resulting in a broader stacked \hi\ profile. The higher star-formation
rate at $z\sim1$ produces greater galactic outflows, of order
$\sim$300\,\kms, which will also affect the measured optical redshifts
(see, for example, \citealt{Weiner2009} and \citealt{Kornei2012}).

We found that when the optical--\hi\ redshift differences are of the same order as
the median \hi\ width, the stacked profile begins to significantly broaden. From the
ALFALFA survey, the median width is $W_{50}\sim150$\,\kms\ at $z=0$, but
this value is unknown at $z>0.5$. Simulations from
\citet{Obreschkow2009} suggest that the size of the \hi\ disk of a
Milky Way type galaxy will decrease by a factor of two between
$z=0-1.5$, with a similar decrease in \hi\ mass. If the
\hi\ disks become significantly smaller, the median $W_{50}$ will also
decrease, and the optical redshifts will need to be of
correspondingly better quality. 

\hi-rich satellite galaxies have not been incorporated in the
current study, and will most likely be undetected in both the optical
and radio data. They may, however, appear in a stacked profile, both in the wings and the
central region of the profile \citep{Khandai2011}. Observationally, at
$z=0$ satellites account for only one~per~cent of the total \hi\ of
the Milky Way, and including the satellites predicted by $\Lambda$CDM simulations,
this increases to $\sim$10~per~cent \citep{Grcevich2009}, and may
become increasingly important at higher redshift where more galaxy
assembly is taking place.

\section{Conclusions}\label{sec:conclusions}

We have conducted an investigation into the relationship between
galaxy redshifts derived from optical spectroscopy and radio \hi\
observations using galaxies from the ALFALFA survey matched to the
SDSS. We find that the redshifts match well, with negligible
offset. The width of the distribution is well approximated with two
Gaussians, the narrow core having a width of 11\,\kms\ and the broad wings
having 36\,\kms. The width of the distribution is narrower if many
optical spectral features are used to measure the redshift instead of
only one feature. The $u-r$ colour was the only optical property of
the galaxies found that correlated with redshift offset, with red galaxies
preferentially lying in the wings of the redshift difference
distribution. Disturbed or interacting systems, as well as galaxies
with asymmetric \hi\ profiles also preferentially have large redshift offsets.

We have constructed simple simulations to investigate the effect
redshift uncertainties have on a stacked \hi\ profile. The simulations
span a range of input mass functions, with the resulting stacked
profiles being similar in each case. In addition to using the velocity
differences derived from the ALFALFA--SDSS galaxies, we also input a
number of increasing velocity differences in order to determine at
which point the redshift errors dominate the properties of the stacked
profile. We find that for redshift errors less than the median width
of the input profiles, the distribution of profile widths is the
dominant effect which broadens the stacked profile. For larger
redshift errors, the input distribution of profile widths becomes
insignificant and the redshift errors broaden the stacked profile.

We have computed the flux for each of the contributing profiles as
well as the flux of the resulting stacked profile, and find that the
recovered flux from the stack matches the average input flux to within
a few per~cent. We have also provided an equation relating the \hi\
mass sensitivity of a stacked profile in terms of telescope and survey
parameters, which will be useful for estimating the number of spectra
required to reach a given average \hi\ mass. This should, however, be treated as an
optimistic case, as the noise in these simulations is Gaussian and may
not be representative of the noise that will be present in real,
interferometric radio data, which will be contaminated by 
RFI and bright continuum sources.

\section*{Acknowledgments}

The Arecibo Observatory is operated by SRI International under a
cooperative agreement with the National Science Foundation
(AST-1100968), and in alliance with Ana G. Méndez-Universidad
Metropolitana, and the Universities Space Research Association. 

Funding for the SDSS and SDSS-II has been provided by the Alfred P. Sloan
Foundation, the Participating Institutions, the National Science Foundation,
the U.S. Department of Energy, the National Aeronautics and Space
Administration, the Japanese Monbukagakusho, the Max Planck Society, and the
Higher Education Funding Council for England. The SDSS Web Site is
http://www.sdss.org/.

The SDSS is managed by the Astrophysical Research Consortium for the
Participating Institutions. The Participating Institutions are the
American Museum of Natural History, Astrophysical Institute Potsdam,
University of Basel, University of Cambridge, Case Western Reserve
University, University of Chicago, Drexel University, Fermilab, the
Institute for Advanced Study, the Japan Participation Group, Johns
Hopkins University, the Joint Institute for Nuclear Astrophysics, the
Kavli Institute for Particle Astrophysics and Cosmology, the Korean
Scientist Group, the Chinese Academy of Sciences (LAMOST), Los Alamos
National Laboratory, the Max-Planck-Institute for Astronomy (MPIA),
the Max-Planck-Institute for Astrophysics (MPA), New Mexico State
University, Ohio State University, University of Pittsburgh,
University of Portsmouth, Princeton University, the United States
Naval Observatory, and the University of Washington. 

We thank the anonymous referee for helpful comments which improved
this paper. NM wishes to acknowledge the South African SKA Project for
funding the postdoctoral fellowship position at the University of Cape
Town. KMH's research has been supported by the South African Research
Chairs Initiative (SARChI) of the Department of Science and Technology
(DST), the Square Kilometre Array South Africa (SKA SA), and the
National Research Foundation (NRF). NM and MJJ thank the South African
NRF for funding a work retreat that contributed to the progress of
this work. We also thank Tom Oosterloo for useful discussions.



\end{document}